\documentclass[transmag]{IEEEtran}

\usepackage{amsmath,amssymb,amsfonts}
\usepackage{algorithmic}
\usepackage{graphicx}
\usepackage{textcomp}
\usepackage[utf8]{inputenc}
\usepackage{amsmath}
\usepackage[style=ieee]{biblatex}
\usepackage[ruled,vlined]{algorithm2e}

\usepackage{latexsym}
\usepackage{amsfonts,amssymb,amsmath}
\usepackage{hyperref}
\usepackage{xcolor}

\def\BibTeX{{\rm B\kern-.05em{\sc i\kern-.025em b}\kern-.08em T\kern-.1667em\lower.7ex\hbox{E}\kern-.125emX}}
\begin{document}

\title{Deep Chaos Synchronization}

\author{Majid Mobini, Georges Kaddoum,
\IEEEmembership{Senior Member, IEEE}
\thanks{This work is supported by the Tier2 Canada research chair entitled ‘Towards a Novel and Intelligent Framework for the Next generations of IoT Networks’.}
\thanks{M. Mobini is with Department of Electrical, Electronics and Communica-
tion Engineering, Babol Noshirvani University of Technology, Babol, Iran.
(email: Mobini2002@gmail.com).}
\thanks{G. Kaddoum is with D´epartement de g´enie ´electrique, University of
Qu´ebec, ´Ecole de technologie sup´erieure, Montr´eal (QC), Canada.(email:
georges.kaddoum@etsmlt.ca).}}

\IEEEtitleabstractindextext{\begin{abstract} \emph{Abstract-} In this study, we address the problem of chaotic synchronization over a noisy channel by introducing a novel Deep Chaos Synchronization (DCS) system using a Convolutional Neural Network (CNN). Conventional Deep Learning (DL) based communication strategies are extremely powerful but training on large data sets is usually a difficult and time-consuming procedure. To tackle this challenge, DCS does not require prior information or large data sets. In addition, we provide a novel Recurrent Neural Network (RNN)-based chaotic synchronization system for comparative analysis. The results show that the proposed DCS architecture is competitive with  RNN-based synchronization in terms of robustness against noise, convergence, and training. Hence, with these features, the DCS scheme will open the door for a new class of modulator schemes and meet the robustness against noise, convergence, and training requirements of the Ultra Reliable Low Latency Communications (URLLC) and Industrial Internet of Things (IIoT).\end{abstract}

}

\maketitle

\section{INTRODUCTION}

\IEEEPARstart{S}{ynchronization}  is a fundamental requirement for a wide range of  natural phenomena and new industrial technologies [1]. 
This paper focuses on designing a new family of chaos-based receivers that benefit from the advantages of Deep Learning (DL) techniques. 
Historically, since the early 1980s, chaos-based communication started with the implementation of electronic circuits exhibiting chaotic behavior [2]. The second fundamental step towards chaos-based communication systems was in 1990. Pecora and Carroll discovered that synchronization can be achieved by coupling two chaotic systems with a common signal [3]. After that, several innovative systems have been proposed employing other outstanding features of chaotic waveforms. These waveforms are wide band, nonlinear, noise-like, non-periodic, and hypersensitive to initial conditions [4]. They also have excellent correlation properties and a simple production process. Thus, chaotic waveforms could be used in a vast spectrum of applications. They are major candidates for multi-user spread-spectrum schemes because of their wideband characteristics [5]- [8]. In addition to the spread-spectrum applications, numerous chaos based modulations have been proposed for use in digital wireless communications because of their robustness against the destructive effects of jamming and fading channels [9]. Moreover, many studies have been done on Low Probability of Interception (LPI) features [10] and chaos-based secure communication schemes [11].\par
In [12], the authors classified secure chaos-based techniques into four generations. The first three generations were based on continuous chaotic synchronization in which the bandwidth efficiency was very low. The fourth generation, called impulsive methods, employed impulsive chaotic synchronization to solve the low bandwidth efficiency drawback. The main design challenge of chaos-based secure schemes is how to deliver a secret message over a public channel using a transceiver structure that shows strong noise rejection capabilities. Several attempts made to design robust digital and analog chaos-based secure communication systems [13], [14]. In general, traditional studies of chaos synchronization rely on the fact that the equations of chaotic systems are known beforehand. This assumption is not realistic for practical chaotic systems in which only one or some observational signals are available [15].\par
 
Many recent studies have demonstrated the capability of ML-based approaches for modeling chaotic systems whose equations of motion are unknown. A special approach known as “reservoir computing” caused considerable progress in model-free prediction of chaotic systems [16], [17]. It is currently an active research area that uses natural potentials of Recurrent Neural Networks (RNNs). Reservoir computing has been widely used in the modeling of temporal processes. Also, reservoir computing has employed successfully in several practical applications, e.g. speech recognition, handwriting recognition, robot motor control, and financial forecasting [18]. More importantly, by transmitting just one scalar signal, synchronization can be obtained between well-trained reservoir computers. This technique can also extract Lyapunov exponents and captures the dynamics of a chaotic system [19].\par
On the other hand, labeling is a difficult and time-consuming process in some practical scenarios, such as chaos-based communications. To tackle this challenge, in our paper we proposed a self-supervised (untrained) structure for a chaos-based receiver. This new design is based on the DIP (Deep Image Prior [20]) approach to achieve a self-supervised configuration. To this aim, we employ a modern version of Generative Adversarial Networks (GANs), known as Deep Convolutional Generative Adversarial Networks (DCGANs), which have a feed-forward structure [21]. Such a design benefits from advantages of conventional DCGANs and self-supervised nature of the DIP approach. Because of these features, the proposed design leads to a high-performance self-supervised structure that relies less on the input data.\par

The results of this paper can be extended to a wide range of applications, from health monitoring, and chaos-based security, to human behavior analysis and pattern studies in dynamic social networks [22]- [25]. The proposed model is flexible for use in a variety of applications with different requirements. It can be used to achieve real-time synchronization in applications that are vulnerable to timing and delay such as control of heart fibrillation. Chaotic conditions of the human heart cause arrhythmia and it can lead to death if left uncontrolled [26]- [28]. To retain a normal heart rhythm, accurate timing can be obtained using the proposed synchronization scheme. On the other hand, there is a lot of evidence that chaotic synchronization provides valuable information and a deeper understanding of physiological mechanisms underlying the brain [29]. When the timing is not the main challenge, de-noising features of the proposed approach can be exploited. Thus, it can be used for noise reduction in practical chaotic signals such as Electroencephalography (EEG), Electromyography (EMG), and other chaotic physiological signals [30]- [32].\par 

\subsection{Basic Definitions}
Synchronisation gives a conception of high correlations between connected systems. In its elementary definition, synchronisation points out to the tendency to have the same dynamical behaviour. In the context of communications, a local copy of chaotic signals can be recreated in the receiver side using appropriate synchronization circuitry. This is the most popular approach in coherent receivers, for recovering the original chaotic samples from the received noisy signals [33]. These recovered samples enters a correlator as a reference signal for data detection. The Bit Error Rate (BER) performance of these receivers depends on the “closeness” of the reference to the original chaotic samples. Using these receivers have some significant advantages over non-coherent receivers in terms of data rate, noise performance, and bandwidth efﬁciency. However, all the above advantages can be attained provided that the synchronization is preserved, Otherwise, all these advantages will be lost [34].\par
To gain a robust chaotic synchronization organization and excellent noise performance, several synchronization policies and various designs for the receivers are reported in the literature. Essential concepts of chaotic synchronization are explained in [3] by finding some effective applications to secure chaos-based communications. In a coupled Lorenz system as a remarkable synchronization scheme [35], [36], two identical chaotic generators are synchronized using a shared drive signal. This type of synchronization is stable and robust to perturbations in the drive signal [37], [38]. \par

In this paper, chaotic map selection and other important design decisions are based on the intended applications. These applications are associated with several important issues such as latency, power consumption, and security [39]. The Lorenz map, as a high dimensional chaotic map, has a more complicated mathematical structure and high chaotic complexity. One-dimensional chaotic maps are more suitable for applications with low latency requirements because they need fewer computational operations. On the other hand, simpler chaotic maps suffer from security limitations. This drawback is due to the limited chaotic range, low chaotic complexity, and higher dynamic behavior degradation rate [40]. In a chaotic masking communication scheme, the robustness to channel noise is a determining measure [41]. In this case, the original message is added to a chaotic signal. At the receiver side, the chaotic signal is subtracted from the received signal to retrieve the original data. In order to mask the original message, this signal should be generally much weaker than the state variable of the transmitter’s chaotic system, and thus the message signal is vulnerable to channel noise. In this situation with high-security requirements, the Lorenz map can be a good choice. In [42] the authors have compared the Lorenz system with a Rössler system and it is shown that in the Lorenz case, even for rather strong noise, the complete synchronization is preserved.\par

In this study, we assume that the received signal is corrupted by Additive White Gaussian Noise (AWGN). This assumption has the following advantages:\par

1) Tractability: Noise effects are very important in chaotic synchronization because of the sensitivity of chaotic systems. In fact, a small noise may lead to instability and synchronization error [43]. This assumption assists in detailed problem tracking by avoiding calculation complexity.\par

2) Necessity: Noise must be considered because it exists always before any other effects of the communication channel. Other destructive effects could be modeled as serial filter blocks after the noise block [44]. \par

3) Generality: The relative performance of different communication schemes determined using the AWGN channel model remains valid under real channel models, i.e., systems under fading conditions generally follow the same trend as in the AWGN channels [44], [45], and [46]. \par

\subsection{Background and Motivations}
To cope with the destructive effects of the noise and distortion, several noise reduction methods have been recommended for chaotic signals, such as the local projection approach and lifting wavelet transform [47], [48]. This issue is even more challenging when the chaotic parameters or initial conditions are unknown. In this case, there are efforts to approximate the noise-free trajectory by estimation of the parameters and initial conditions [49]- [51]. However, in most of the reported works, the chaotic signal cannot be recovered precisely. On the other hand, since noise and chaos have similar spectral behaviour, regular frequency domain filtering degrades the signal of interest [52]. \par
Recently, DL-based techniques have attracted great attention because of effectiveness in data-driven analysis of nonlinear dynamics [53]-[56]. Among these, Recurrent Neural Network (RNN) is frequently used to learn complex dynamics from the input data [57]-[60]. Unfortunately, classical RNN methods suffer from vanishing gradients and tendency to take into account only short-term dependencies [61]. In [62], the authors compared the prediction performance of the RNN and Long-Short Term Memory (LSTM) in the presence of noise. They showed that when LSTM was trained on noisy data, it reduced the noise contribution in the prediction process. This feature can be used for noise reduction goals. However, when LSTM was trained on clean data it became susceptible to perturbations in the input data.\par

In a surprising new article [20], Deep Image Prior (DIP) method is suggested for image de-noising and inpainting using  DCGANs [21]. In DIP, a deep convolutional neural network generator such as DCGAN is initialized with some random weights and these weights should be optimized in such a way that force the network to generate an image that resembles the target image as much as possible. This procedure does not use any prior information, or training on large data sets. Motivated by the potentials of the DIP method [20], this article presents some approaches to take advantages of the DL in order to promote the capabilities of the existing chaos-based communication systems.
\subsection{Contributions }
This paper is the first to introduce DL-based chaotic synchronization along with some approaches to improve trainable and untrained systems. In chaotic synchronization and other communication systems, large data sets and labeled data are not available. In addition, for a reasonable performance in regular DNN-based structures, learning and testing steps must be performed for the same channel realization results in a lot of overhead due to learning [63]. Innovative aspects of this paper are briefly itemized below:\par
$\bullet$ As the main achievement, we introduce a Deep Chaos Synchronization system (DCS) that needs no data set.
The proposed DCS benefits from the inherent security of the chaotic signals and advantages of DNNs. Moreover, it can offer a low synchronization error, especially for longer sequences. The above-mentioned features make DCS an appealing candidate for use in chaos-based Code Division Multiple Access (CDMA) systems [64], encrypted data transmission [65], [66], Ultra-Reliable Low Latency Communications (URLLC) [67], Industrial Internet of Things (IIoT) [67], and Wireless Sensor Networks (WSNs) [68], [69].\par
$ \bullet$~In addition, we suggest a trainable RNN-based synchronization structure  gains de-noising properties of the sequence to sequence RNN, as a special arrangement of the RNNs family for comparison with our proposed DCS approach. We accepted the traditional RNN as a well-known trainable approach since it can give a measure for the de-noising strength of the proposed method. Depending on the intended application, each has its own benefits and can be used.\par
In order to initial condition estimation, we offer a Genetic Algorithm (GA)-based approach, which gives a suitable reconstruction of the transmitted signals. A correct estimation is vital in chaotic synchronization because of their sensitivity to the initial condition. With accurate initial conditions, the slave system can be synchronized with the master system almost instantaneously, thereby eliminating the problem of transients. This technique can still keep synchronization where the largest conditional Lyapunov exponent is positive and other strategies cannot be synchronized [35].\par
In order to select a proper map among various chaotic maps, we tested several chaotic maps as the DCS system input. We found a tractable trade-off between processing time and noise robustness. In other words, we show that using the Rössler map [40] and discrete-time Henon map [72], the processing time reduced. The average amplitude error between the recovered drive signal and the original signal shows that the Lorenz map is superior to other maps from a noise robustness point of view.\par

The rest of this paper is organized as follows: In Section II, the basics of the Lorenz map and some other chaotic maps are investigated. Also, the employed initial condition estimation method is described in this Section. Moreover, some basic information about DCGANs, the structure of the proposed transmitter/receiver, and other information related to the DCS system is presented in Section II. In Section III, the RNN-based synchronization system is introduced for comparison with our DCS method. In Section IV, simulation results and discussions are presented. The conclusions are explained in Section V. \par

\begin{figure}[htp]
    \centering \includegraphics[width=9cm]{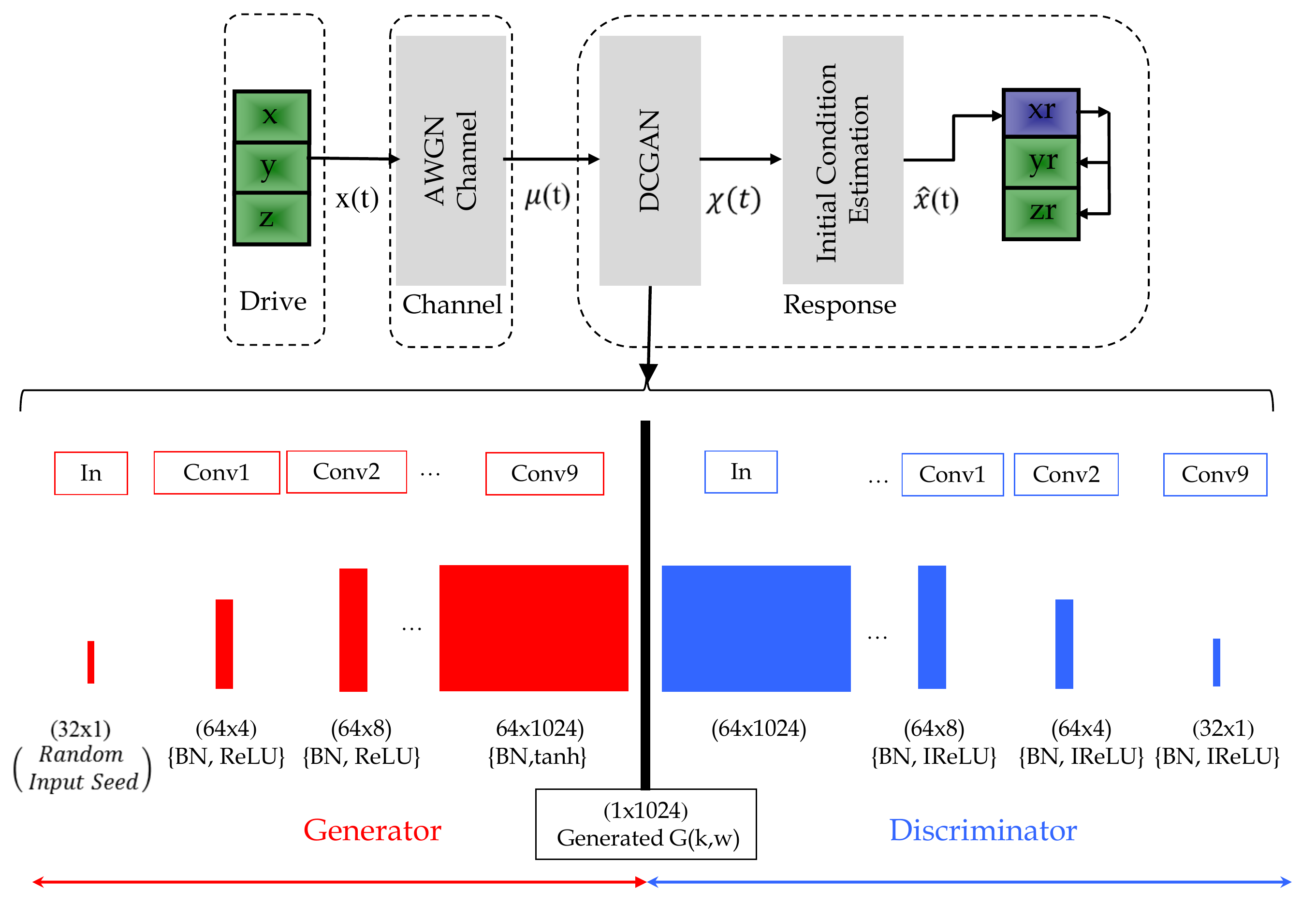}
    \caption{DCS model based on DCGAN as the generative model.\label{fig1}}   
    \label{fig:Picture1}
\end{figure}

\section{DCS MODEL}
In this section, the basics of the Lorenz chaotic synchronization system are revisited and then the offered Deep Chaos Synchronization model is presented.

\subsection{Conventional Lorenz System}

Consider the following conventional Lorenz system of Equations [1]:
\begin{equation}\label{eq1}
\centering
\begin{split}
 \dot x= \rho(y-x), \\
 \dot y= rx-y-xz,\\
 \dot z=xy- \beta z, \\
 \end{split}
\end{equation}
where $x$, $y$, and $z$ are the state variables at the drive (Master) subsystem and $\rho$, $r$, and $\beta$ are control parameters. When these parameters are chosen as $\rho=10$, $r=28$, and $\beta=8/3$ the equations lead to chaotic dynamics while the initial conditions shape the chaotic behavior [3].
At the receiver side, the received noisy signal can be written as:\\
\begin{equation}\label{eq2}
\centering
 \mu (t)=x(t)+\zeta (t),
\end{equation}
where $\mu (t)$ is a given noisy observation, $\zeta (t) $ is Additive White Gaussian Noise with zero-mean and power spectral density $\sigma_n^2$. The response (Slave) uses the $\mu (t)$ as driver, in order to reconstruct Lorenz state variables $y_r$ and $z_r$:\\
\begin{equation}\label{eq3}
\centering
\begin{split}
 \dot y_r= r\mu -y_r- \mu z_r,\\
 \dot z_r= \mu y_r- \beta z_r. \\
\end{split}
\end{equation}\par
We presume that all control parameters $\rho$, $r$, and $\beta$ of the master sub-system are known at the receiver side. The initial values of the master subsystem are $(x_0, y_0, z_0)$ and the slave subsystem is initiated similarly, except that we do not know one of the initial values $(x_0)$. In a Lorenz system starting from two slightly different initial states $(x_0, y_0, z_0)$ and $(x_r{}_0, y_r{}_0, z_r{}_0)$, the trajectories diverge from each other. At the end of this section, a GA-based solution to this challenge is presented.

\subsection{Comparison with Other Chaotic Maps}

In addition to the Lorenz system, there are many other chaotic systems that generate chaos and can be utilized in communication systems. For example, Rössler systems can be decomposed into stable drive-response sub-systems to produce a robust synchronization [40]. Moreover, several discrete-time maps, such as the Henon map, are also decomposable into synchronizing sub-systems without the burden of solving differential equations [72]. In Section IV, we discuss the merits and demerits of using these systems. To investigate the effect of different maps on de-noising performance of the DCS system, we first adopt a signal from the Rössler system as input [40]:
\begin{equation}\label{eq1}
\centering
\begin{split}
 \dot x= \omega y-z, \\
 \dot y= \omega x+0.15y,\\
 \dot z=0.4+z(x-8.5),\\
 \end{split}
\end{equation}
with $\omega=0.95$, and another chaotic signal from discrete-time Henon system [72]:
\begin{equation}\label{eq1}
\centering
\begin{split}
 x[n+1]=1+y[n]-z[n]x^2[n], \\
y[n+1]=bx[n],\\
z[n+1]=z[n]-0.5+\beta x^2[n],\\
 \end{split}
\end{equation}
with $b=0.25$, and $\beta= 0.279$.\par

\subsection{Proposed DCGAN-based Receiver}

In this study, motivated by the appealing factors of the DIP model [20], the one-dimensional arrangement of DIP is implemented using a DCGAN. In particular, we adapt the input and output of the DIP model and use it as a baseline which we term the Deep Chaos Synchronization. The receiver of the DCS is composed of three stages, as shown in Fig. 1. The implementation steps of the DCS are illustrated in Algorithm 1.

 In the first stage, a less noisy signal is achieved from the DIP method. When the objective is to generate data identical to a target distribution, a Generative Adversarial Network (GAN) provides data from a similar distribution efficiently [70]. However, GANs are unstable to train and this regularly generates nonsensical outputs. 
 
DCGAN is one of the attempts to scale up GANs using CNNs. Compared to conventional GANs, spatial pooling functions are replaced with convolutional layers, allowing the DCGAN to learn its own spatial downsampling. Also, DCGAN adopts Batch Normalization (BN) which stabilizes learning by normalizing the input to each unit to have zero mean and unit variance. Moreover, ReLU activation functions are considered for all layers. The RMSProp optimizer is used as a variant of the traditional gradient descent method that adopts a momentum parameter for faster convergence of the training process.

In the second stage, we offer a GA-based approach in order to estimate the initial condition using the achieved clear signal from the previous stage. An accurate initial state estimation results in a proper reconstruction of the original signal. 

In the final stage, the estimated initial value enters the response sub-system in order to reconstruct Lorenz attractors. By doing the above mentioned three stages the inherent advantages of a chaotic system are combined with the noise reduction capability of the DIP so as to have remarkable performance. However, the complexity increases in comparison with the conventional Lorenz synchronization system, but compared to RNN-based designs, DCS needs no labeled data or large data set. Moreover, no need to send train signals result in a lot of overhead reduction.

As shown in Fig. 1, the DCS receiver is implemented using a DCGAN located at the input of the response subsystem. We aim to tackle the inverse problem of de-noising a drive signal $x(t)\in \mathbb{R}^N$ which has been transmitted on a communication channel with Gaussian noise $\zeta (t)\in \mathbb{R}^N$. Therefore, given a noisy observation $\mu(t)$, the basic idea is to reproduce a signal $ \chi (t) $ that is similar to $ x (t) $. To implement the DCGAN, We need to configure and organize the three blocks: 1) the generator, 2) the discriminator, and 3) the training process. The  discriminator can be implemented  by successive convolutional layers, BN layers, and Relu activation functions. In each of the convolutional layers, we downsample the spatial dimension of the input. The generator of the DCGAN consists of a sequence of transpose convolutional layers that upsample the input random sample to generate an artificial signal.
 A DCGAN can be trained in the same way as a regular GAN. Different from the conventional DCGAN models, in the DIP approach, we optimize the "$random~weights$" to force the network to produce an output similar to the target. Therefore, G (k, w)is the output of DCGAN with weights $w$ and latent vector $k$. In other words, instead of minimizing $ \|\chi -\mu\|^2$, the network optimization problem of the DCS can be formulated as:
\begin{equation}\label{eq3}
\centering
w^*=\underset{w}{\mathrm{argmin}}\|G(k,w)-\mu\|^2,
\end{equation}
where $ \mu\in \mathbb{R}^N $ is the observed noisy signal and $G(k,w^*)=\chi$ is the output clean sequence. This non-convex optimization problem can be solved easily using Gradient Descend (GD) method. As an alternative optimization strategy, we can regularize the network using a two stages optimization, which has shown converges better than simple Mean Squared Error (MSE) optimization [71], [73]. Therefore, we first train the adversarial network using a noisy sequence $\mu$, an RMSProp optimizer, and the loss function in (6). Then, this problem can be solved through a simple GD algorithm to obtain the final solution.\\

\begin{algorithm}
\SetAlgoLined

1: ~~~~Initialize:\\~~~~~~~Observed Noisy Sequence $\mu$, Random vector $w$;\\
2: ~~~~DCGAN pseudo-code:\:\\
~~~~~~~~2-1: Generator;\\
  ~~~~~~~~~~~~~~- 1D Convolution Transpose \\
  ~~~~~~~~~~~~~~- Batch Normalization \\
 ~~~~~~~~~~~~~~- Relu \\
 ~~~~~~~~~~~~~~- Tanh ~~ (At the end of Layers)\\  
 ~~~~~~~ 2-2: Discriminator (Input random vector k)\\
   ~~~~~~~~~~~~~~- 1D Convolution transpose\\
   ~~~~~~~~~~~~~~- Batch Normalization \\
  ~~~~~~~~~~~~~~- IRelu \\
~~~~~~~ 2-3: DCGAN training loop \\
  ~~~~~~~~~~~~~~- Draw  training example from the $\mu $ and $w$\\
   ~~~~~~~~~~~~~~- Generate a network $G(w,k)$\\
   ~~~~~~~~~~~~~~- Network training using (6)\\
   ~~~~~~~~~~~~~~~~$w= RMSprop (G(w,k), LR, MOM)$\\
    ~~~~~~~~~~~~~~- Update loss\\        
    ~~~~~~~~~~~~~~~ $Output=Network(w)$\\

   3: ~~~~DIP Output\\
       ~~~~~~~~~~~~~~- Solving (6) using GD method\\
                ~~~~~~~~~~~~~~~ $ \chi = G^*(w,k)$\\

    4: ~~~~Initial condition estimation $(\chi, \hat x)$\\
~~~~~~ ~~~~~~~-~GA based calculation of the $x_0$\\
 5: ~~~~Reconstruction of  $x_r, y_r, z_r$\\
 6: ~~~~End\\
 \caption{ DCS Receiver Pseudo-code}
\end{algorithm}

\subsection{Initial Condition Estimation using GA}\par
The chaotic trajectory reproduction is extremely susceptible to the initial conditions. On the other hand, in most practical scenarios, the initial conditions are hard to measure precisely. So, as a reasonable assumption, we assume that we have only the upper and lower bounds for initial conditions. After an effectively initial condition estimation, we can get a noise-free signal, and the slave system can follow the dynamics of the master system almost instantaneously, thereby eliminating the issue of transients. With these advantages, initial condition estimation is one of the hot topics in chaos-based studies. Some articles have emphasized that when alternative synchronization techniques fail, this approach can achieve synchronization [74-78].  For example, in [48], [75] some heuristic algorithms are employed to initial state estimation and
robustness of the chaotic synchronization to noise.\par
In this paper, in order to initial condition estimation, we offer a simple GA-based method that provides an excellent reconstruction of the transmitted signal. The GA-based optimization is a convenient tool for non-linear problems that are not appropriate for analytical optimization methods. In addition, GA approach is fit for solving complex multi-variable problems, for example, where we would like to estimate several chaotic parameters and initial conditions. In the literature different methods have been proposed for initial condition estimation, e.g. Differential evolution algorithm-based methods [58], or Hybrid Mehta-heuristic methods [59]. Since the main idea behind this study is to implement an untrained structure for chaos-based receivers and our main focus is not on the precise parameters and initial value estimation, we select the GA to simplify the implementation.\par

We focus on the simultaneous start property and the security aspects of the initial condition estimation method [73]. The simultaneous start property is important in the synchronization of two coupled discrete-time systems, because it enables us to obtain real-time synchronization without transient time. The receiver can start de-noising process after receiving the first sample. In other words, because the receiver gains deep learning structures, it is inherently predictive, and after receiving the first sample, it can predict the next samples. In the synchronization model presented, if the response sub-system knows the initial conditions of the transmitter, or estimates them accurately, it can make a copy of the transmitted signal without the need for additional signaling. 
From a security point of view, when the adversary network has access to the wireless medium and public samples $x_0$, we can secure the DCS system by sharing $y_0$ and $z_0$  with legal users as private keys. We can use any of the transmitters outputs $ x_0,$ $y_0$, and $z_0 $  for signal transmission or security purposes. 
Estimating other variables imposes time and computational load and without loss of generality, we take a single variable case $x_0$ to decrease the computational complexity.\par
In our setting, equal parameters $\rho$, $r$, and $\beta$ are considered for the drive and response systems because the synchronization between them is stable provided that the parameters are identical.
Here, $ (x_0, y_0, z_0) $ and $ (x_r{}_0, y_r{}_0, z_r{}_0) $ are the initial states of the drive and response systems, respectively. We have adopted the $x$ component to operate as a drive signal and it is also assumed that the receiver does not know the initial state of the $x$ component (I. e. $ x_0$ ) correctly and only knows the upper and lower bounds of it. To cope with the above-mentioned  trajectory divergence phenomenon, in [75], the authors recommended a piece-wise estimation for each segment of the observed signal. Mathematically, initial state estimation  in such a problem reduces to the optimization of the following objective function:\\
\begin{equation}\label{eq3}
\centering
\underset{x_0}{\mathrm{argmin}}
{[f_o{}_b{}_j (x_0)= \frac{1}{T_s}\sum_{t=1}^{T_s}{(\hat x (t) - \chi (t))}^2], }
\end{equation}
where $x_0$ is initial condition of the $x$ component, for the Lorenz map. The idea is to minimize
the objective function $f{}_o{}_b{}_j (x_0) $ given by the mean-squared error between a candidate chaotic signal solution $\hat x(t)$ and the output of the above mentioned neural network $\chi (t)$. If an initial population is determined for $(x_0)$, the solution will be found by an iterative approach such as GA. The initial population can be modified based on some rules such as selection, mutation, and crossover to achieve the optimal solution [79]. In particular, we use a decimal representation of genes in the range [0 0.1], uniform mutation, and one-point crossover. It is necessary to notice that here we determine the accuracy of our estimates by the initial population definition. In other words, the larger the initial population, the higher the accuracy of estimation, and accordingly the reconstructed signal $\hat x$ will be closer to the primary signal $x$. Iterative solving of the system of equations in (1) can give the best solution $\hat x (t) $. This can be performed in Python with the help of \texttt{scipy.integrate.odeint}
 function. Ultimately, we can reconstruct Lorenz attractors $y_r$ and $z_r$ at the receiver side using parameters and the estimated initial condition.

\section{TRAINABLE RNN-BASED SYNCHRONIZATION MODEL}
\label{sec:guidelines}
In this section, we introduce a trainable RNN-based synchronization scheme for comparative investigation. 
Fig. 2 illustrates the suggested model.

\begin{figure}[htp]
    \centering \includegraphics[width=9cm]{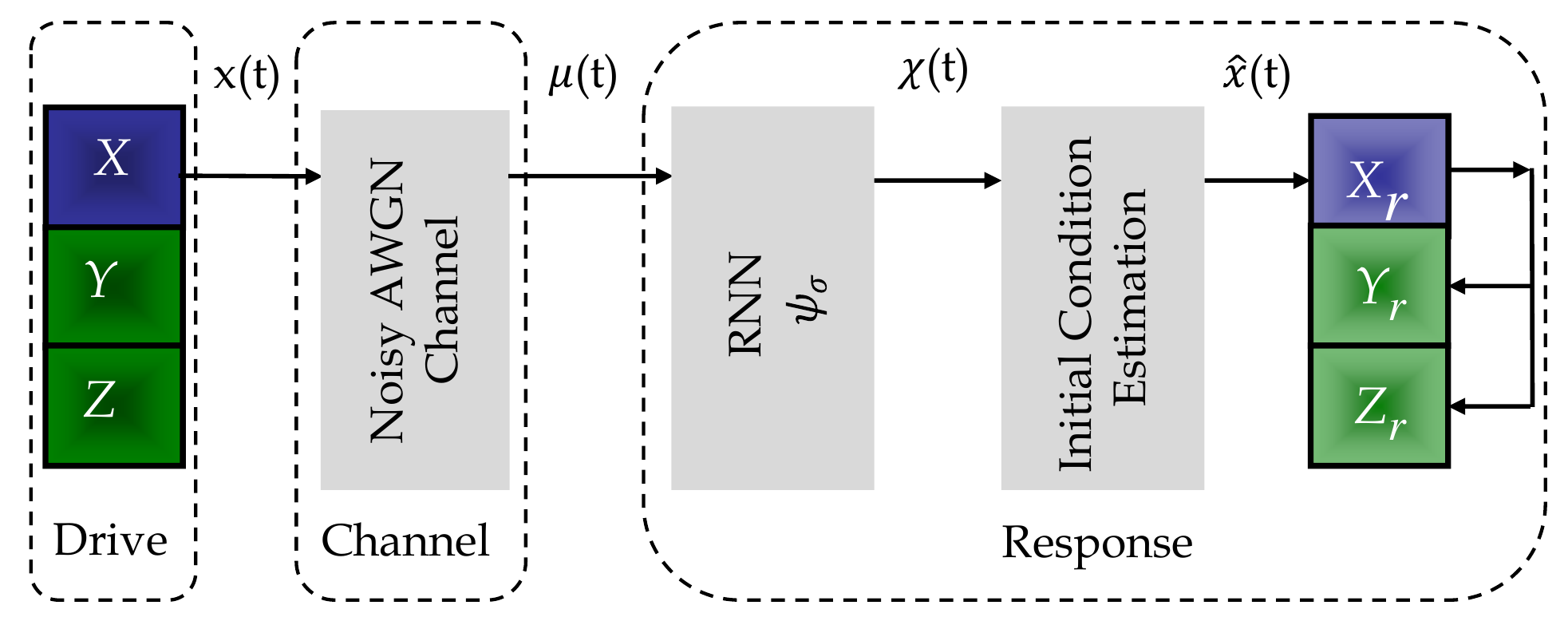}
    \caption{Trainable RNN-based synchronization System.\label{fig1}}   
    \label{fig:Picture1}
\end{figure}

It is noted that since the DCS is the ﬁrst self-supervised chaotic synchronization system, the performance comparison cannot be provided with other self- supervised systems. Therefore, the comparison with the RNN-based synchronization system can provide a general measure of de-noising power of them, although, they belong to the two different training classes.\par

A standard RNN includes of a hidden layer $\bf h$ and an output $\bf \chi$ works on an input sequence $\bf x$= ${x(1), . . . , x(T)}$. The hidden state $h(t)$ is updated at each time step as follows:\\
 \begin{equation}\label{eq5}
\centering
\ h(t+1)=\psi (x(t), h(t)), 
\end{equation}
where $\psi$ is a non-linear activation function, such as a logistic sigmoid function. An RNN learns a probability distribution over a sequence by being trained to predict the next symbol in a sequence. The output at each time step $t$ is the conditional distribution $p(x(t+1)|x(t), . . . , x(1))$. From this learned distribution, it can sample a different sequence by iteratively sampling. 
The RNN is a remarkable tool for learning complex dynamics and, at the same time, has a lot of noise reduction capabilities. It is also proper for Natural Language Processing (NLP) and signal prediction problems. Among these capabilities, we focus on the de-noising potentials of the Encoder-Decoder sequence to sequence RNNs as a special structure of the RNNs family. It consists of two RNNs, one of them encodes a sequence into a fixed-length vector and the second RNN decodes the vector into another sequence [80]. 
In the RNN-based signal prediction problem,  the model predicts the next sequence given a previous sequence of vectors. This will enable us to generate a new sequence, one vector at a time. The architecture of the de-noising problem is similar to the prediction process, except that the signal given to the input of encoder is noisy, and the desired output of the decoder should be a clear signal.  We adopt a newer version of dynamic RNN functions, known as Gated Recurrent Unit (GRU), which realizes the RNN for us [81]. GRU has no output gate and incorporates the input and previous gates into an update gate. This new gate manages how much the internal state is combined with a candidate activation.\par
We select GRU because it requires less computation while provides comparable results with LSTM. In addition, GRU can be easily implemented by \texttt{GRUCell} in TensorFlow. Different GRU structures are supported by TensorFlow in the \texttt{tf.nn.rnn\_cell} module [82], [83]. Fig.3 illustrates the arrangement of the stacked GRU sequence to sequence RNN.
\begin{figure}[htp]
    \centering \includegraphics[width=9cm]{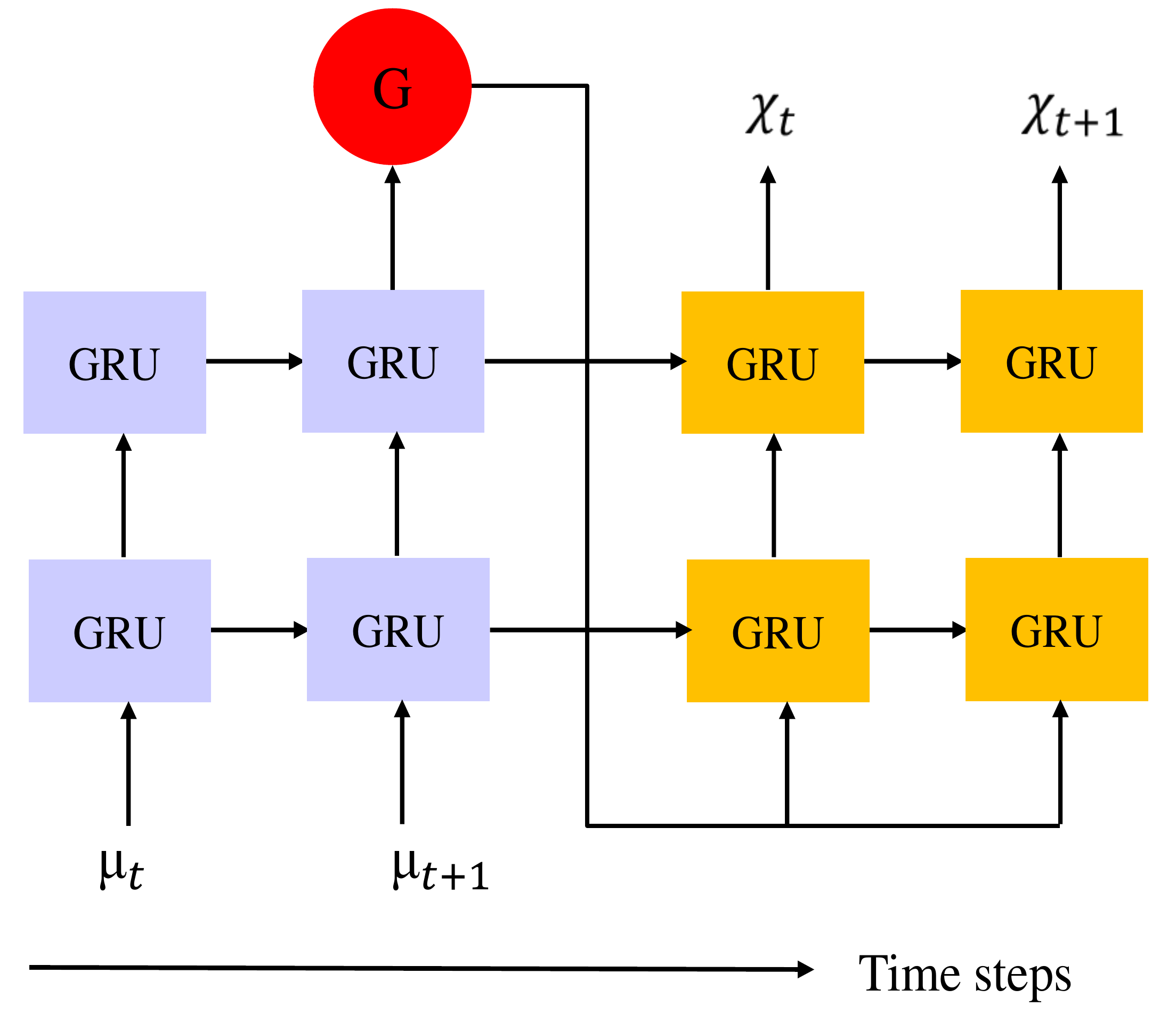}
    \caption{The stacked GRU sequence to sequence RNN structure.\label{fig1}}   
    \label{fig:Picture1}
\end{figure}

The $G$ is the value of the hidden output of the encoder in the last time step. We should repeat this value as an input to the further RNN to make it remember the context of the present at all moments when predicting the future. 
We have now defined the RNN and should train the model using an optimizer. 
The input and output neuron’s dimensions should be $m=d=1$ because the number of input train and test sequences are one $n{}_t{}_r{}_a{}_i{}_n=n{}_t{}_e{}_s{}_t=1$. We consider a supervised framework and generate two sequences of the noisy chaotic samples with length $T$ as $train$ and $test$ signals. In the meantime, we keep their noise-free version (or $target$) for final evaluation. In this framework, we train the model showing the $train$ sequence and reading a prediction from the output neurons in the test phase.
Even with a corrupted $test$ signal, this prediction is close to the ground truth [59]. The RNN is trained to suppress the contributions from the noisy observations and minimize the following objective function using an ADAM optimizer [84]:
\begin{equation}\label{eq5}
\centering
 L=\sum_{t=1}^{N} \frac{1}{2}(\mu(t)-\chi(t))^2 .
\end{equation}

Here,\\
\begin{equation}\label{eq5}
\centering
\chi(t+1)=\psi_\sigma (\mu(t+1), G(t)), 
\end{equation}
where $h(t)\in \mathbb{R}^N$ are the internal states, $N$ is the number of memory units, and $\chi(t)$ is the predicted/de-noised signal. $\psi_\sigma$ denotes the RNN trained against noise with a standard deviation $\sigma$. In order to reconstruct Lorenz attractors, $\chi$ enters the initial condition estimation block. 
After initial state estimation, the desired attractors can easily re-produced by the Lorenz circuit.
\section{SIMULATION RESULTS}
In this section,  a number of simulations are considered for a comparative analysis of the DCS, RNN-based synchronization, and traditional Lorenz coupled system in (3). The considered parameters are listed in Table I.
\begin{table}[!t]
\renewcommand{\arraystretch}{1.3}
\small{\caption{\small{List of the Parameters.}}}
\centering
\begin{tabular}{|c||c||c||c|}
\hline
$~Variable~$ &$~~~Description~~~$ & $~RNN~$ & $~DCS~$\\
\hline
N & Size of the RNN & 64 & -\\
\hline
$n{}_t{}_e{}_s{}_t$  & Number of test signals & 1 & 1\\
\hline
$n{}_t{}_r{}_a{}_i{}_n$ & Number of training signals & 1 & -\\
\hline
T & Length of train / test signals & 1024 &1024\\
\hline
$N{}_G{}_F$ & Convolutional filter per layer & $-$ & 64\\
\hline
Iter & Iterations during the training & 800 & 800\\
\hline
Mom & Momentum of RMSProp & $-$ & 0.9\\
\hline
m & Output dimension & 1 & 1\\
\hline
d & Input dimension & 1 & 1\\
\hline
LR & Learning Rate & $10^-{}4$ & $10^-{}4$ \\

\hline
\end{tabular}
\end{table}
Many methods have been proposed to implement chaotic trajectories in python. We solve the Lorenz equations, using 
\texttt{scipy.integrate.odeint}
function according to the method presented in [85]. This function integrates the system of Ordinary Differential Equations (ODE) and returns the solution. We produce 1024 data points with a time step= 0.1.
It is presumed that all control parameters of the master sub-system are known at the receiver side. To produce chaos dynamics, the control parameters are selected as $\rho=10$, $r=28$, and $\beta=8/3$. The initial values of the master subsystem are $( x_0, y_0, z_0)=(0.1, 0.1, 0.1)$ and the slave subsystem is the same, except for one of the values $(x_0) $, which is assumed to be unknown and  must be estimated. To this aim, we offer the GA-based initial condition estimation method. The parameters of the GA-based approach are listed in Table II. 

\begin{table}[!t]
\renewcommand{\arraystretch}{1.3}
\small{\caption{\small{GA Parameters.}}}
\centering
\begin{tabular}{|c|}
\hline
$Variable
~~~~~~~~~~~~~~~~~~~~~~~~~~~~~~~~~~~~~~~~~~~~~~~Description$\\
\hline
population size
~~~~~~~~~~~~~~~~~~~~~~~~~~~~~~~~~~~~~~~~~~~~~10000\\
\hline
Number of variables ~~~~~~~~~~~~~~~~~~~~~~~~~~~~~~~~~~~~~~~~~~~~~1\\
\hline
Maximum iterations ~~~~~~~~~~~~~~~~~~~~~~~~~~~~~~~~~~~~~~~~~~~~10\\
\hline
Crossover  fraction
~~~~~~~~~~~~~~~~~~~~~~~~~~~~~~~~~~~~~~~~~~~~~0.1\\
\hline
Mutation fraction
~~~~~~~~~~~~~~~~~~~~~~~~~~~~~~~~~~~~~~~~~~~~~0.1\\
\hline
\end{tabular}
\end{table}

\subsection{De-noising Performance }
We can easily make the RNN model using NumPy and TensorFlow packages to produce a clear signal from a noisy observation. This ability comes from that the RNN tends to weaken the contribution of the noise and captures its own learned dynamics. Fig. 4 shows the convergence curve of the RNN, in which the loss is plotted versus the number of iterations during the training process.  In this simulation, the RNN is trained at $\sigma^2_n=0.3$ while is tested at $\sigma^2_n=0.5$. If $\sigma^2_n$ is set at a big value, the RNN might learn nothing but the noise. On the other hand, if the training $\sigma^2_n$ is low, the RNN can only learn the clean signal. For a  proper training $\sigma^2_n$  value must help the RNN to learn both types of samples.
The proposed RNN-based system can be
trained at a speciﬁc and proper noise condition, while operating throughout the range. during the training stage, we employ ADAM optimizer to minimize the loss function presented in (9). One can see that the RNN converges to the best case after several iterations. This demonstrates that the training phase not only imposes computational complexity, but also can be time consuming for the receiver and results in a lot of overhead.
\begin{figure}[htp]
  \includegraphics[width=8.5cm]{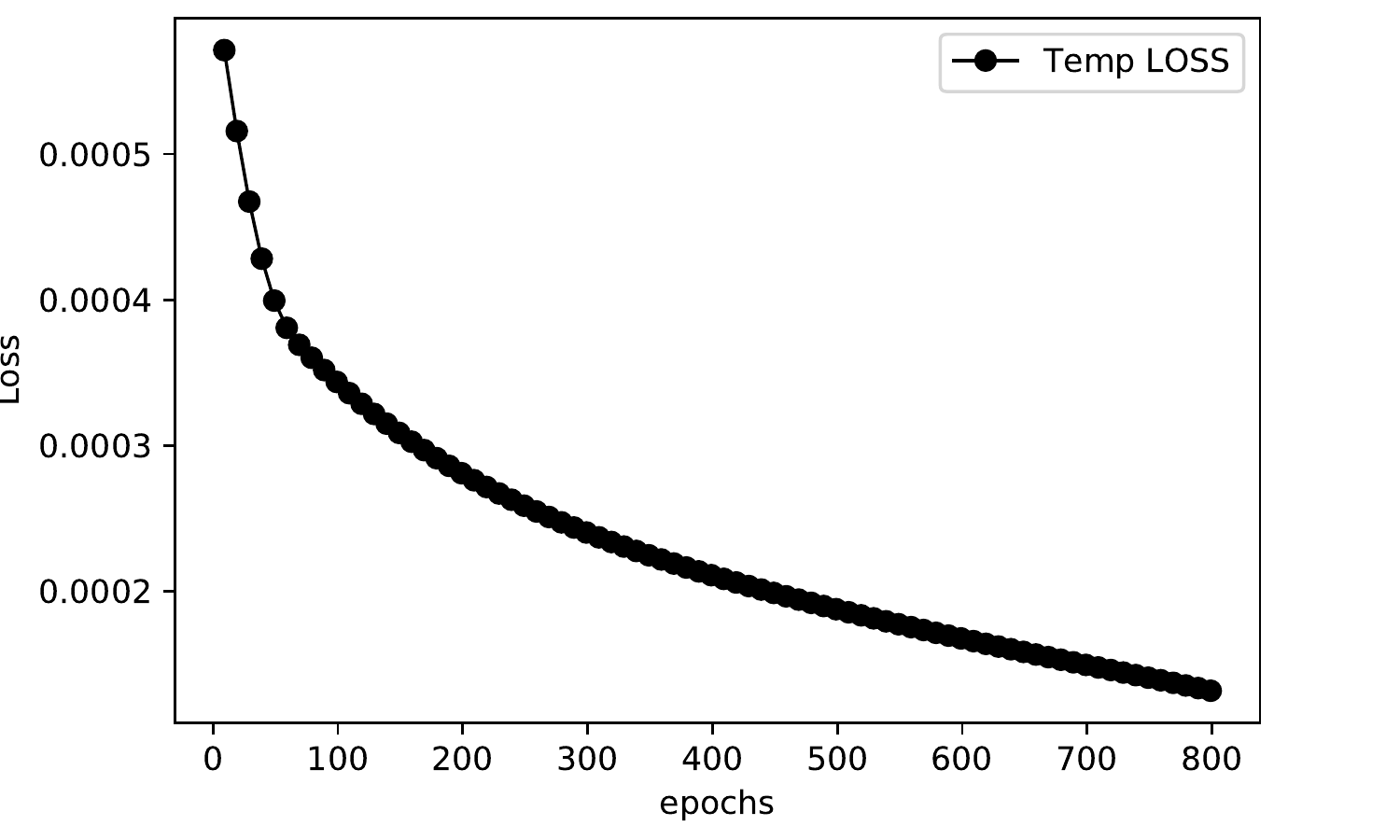}
  \centering
    \caption{\small {Optimizing iterations during the training of RNN.}}
\end{figure}\par
Fig. 5 (a) compares the target $(x)$ and de-noised $(\chi)$ signals. In order to facilitate visual comparison, the first 250 samples of these signals are plotted. The RNN model follows the target dynamics, but the synchronization is degraded after about 200 time steps. In other words, the RNN is not able to capture the dynamics of the drive signal and its error signal has a large domain. Therefore, it is necessary to re-train the receiver after the above time. Re-training process consumes time and other resources, and results in high computational complexity. In contrast, as shown in Fig. 5 (b), the DCS model, which is implemented using Torch and TensorFlow libraries, follows the driving dynamics perfectly.
\begin{figure}

\includegraphics[width=8.5cm]{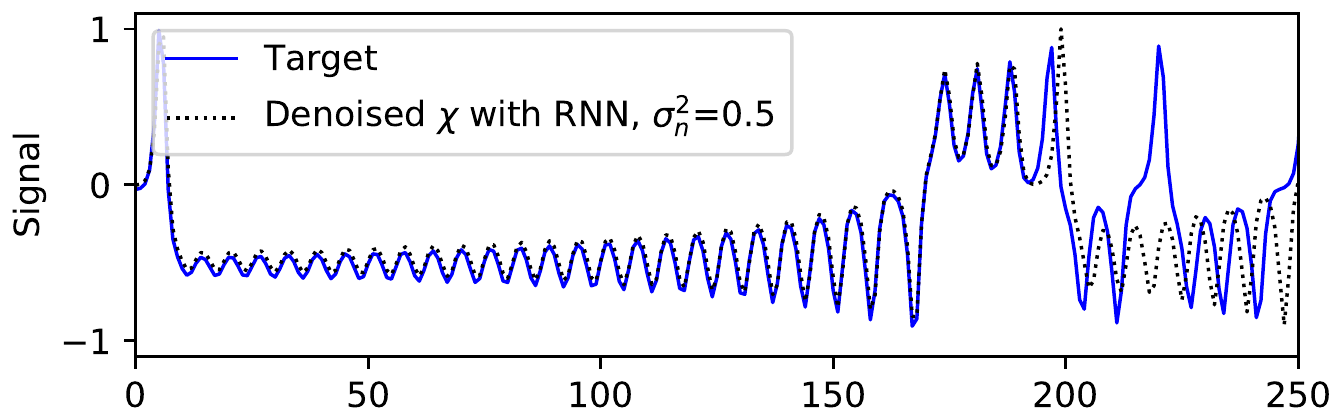} 
\begin{center}
t\\
(a)
\end{center}

\includegraphics[width=8.5cm]{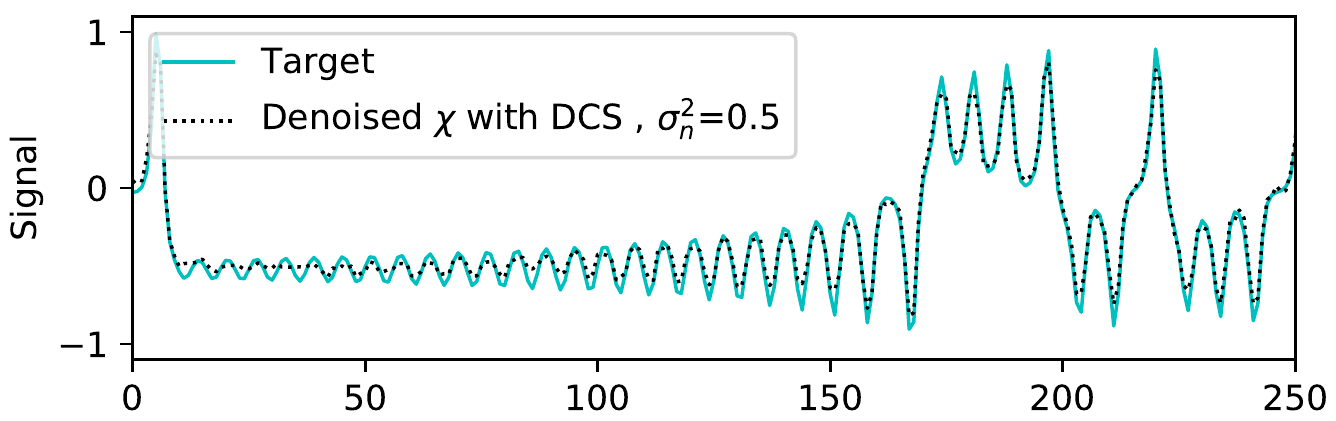}
\begin{center}
t\\
(b)
\end{center}
\caption{\small {(a) Target and de-noised signal ($\chi $) by RNN (b) Target and de-noised signal ($\chi $) by DCS.}}

\end{figure}\par
Fig. 6 shows de-noising error, i.e., $(x(t)-\chi(t))$ of the DCS compared with the RNN system in the same scale. We observe that, although the DCS does not use pre-training, it follows dynamics of the target signal and the error amplitude remains fixed over time.
 \begin{figure}[htp]
     \includegraphics[width=8.5cm]{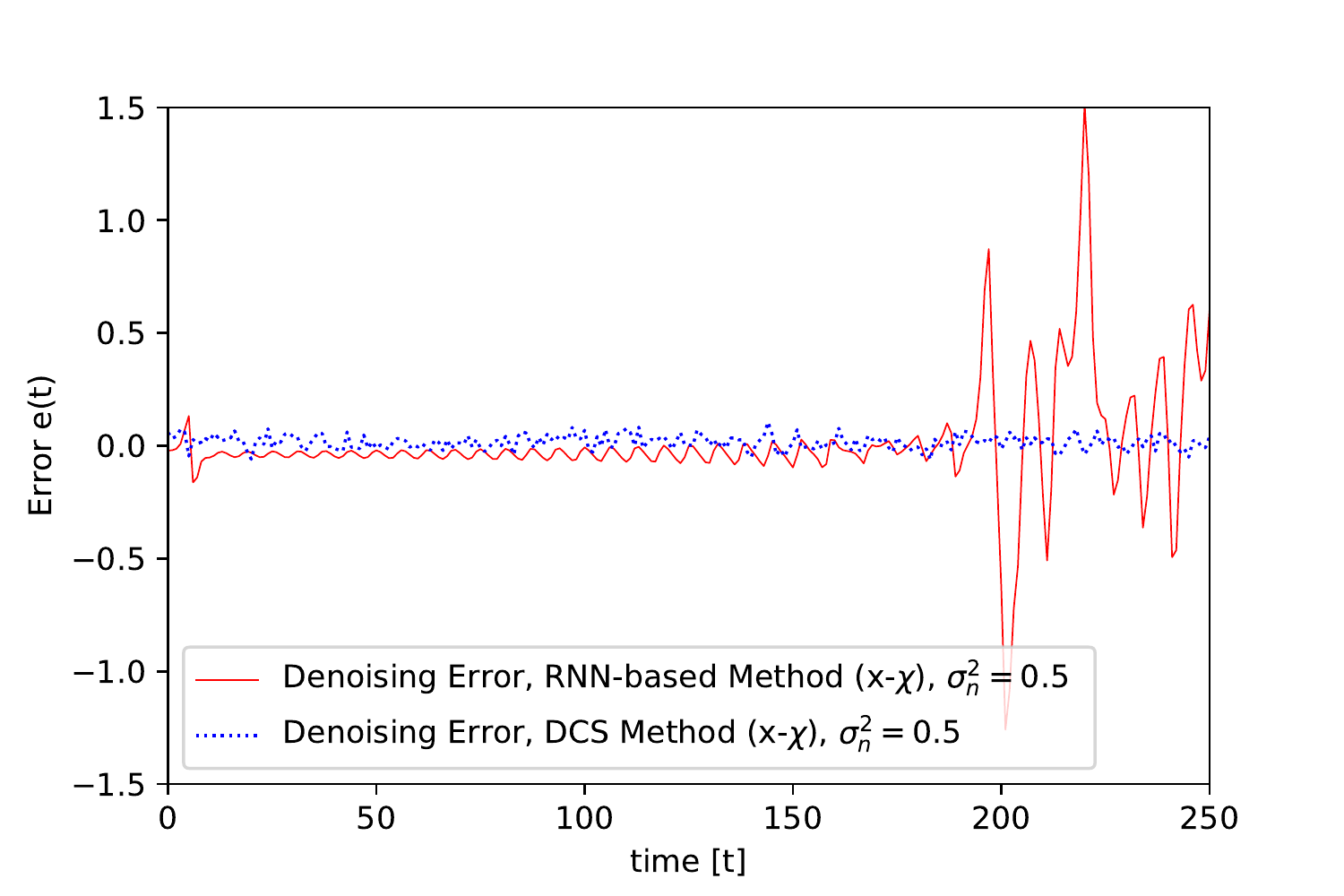}
    \caption{\small { Comparison between de-nosing error $x(t)-\chi(t)$ of the RNN and DCS.}}
    \label{fig:Picture1}
\end{figure}
Fig. 7 (a) plots the average amplitude error of the DCS for different number of iterations. The average amplitude error converges to the best case rapidly after about 600 iterations. Fig. 7 (b) shows the average amplitude error between the recovered drive signal and the original signal for different values of $\sigma_n^2$ . As illustrated in this figure, for large noise values, when the number of iterations increases to 800, the error  becomes very small. Increase the number of iterations from 800 to 2000 has a negligible impact on the error performance.
In Fig. 7 (c), the robustness of the DCS system is measured for different chaotic maps. Execution time of the DCS algorithm is also measured for each of the maps. The average amplitude error between the recovered signal and the original signal shows that even for a rather strong noise, the Lorenz map is more robust. However, when we use the Rössler and Henon maps presented in (4) and (5), the processing time is reduced by $20\% $ and $25\%$, respectively. The Lorenz map has a more complicated structure and high chaotic complexity, while the Rössler and Henon maps are faster and more suitable for applications with low latency requirements such as URLLC and IIoT. On the other hand, simpler chaotic maps suffer from security limitations, and the Lorenz map is more suitable for applications such as synchronization in extremely noisy channels and secure communications. According to the above discussion, future systems should make different compromises to meet their requirements.

\begin{figure}[h]

\includegraphics[width=8.5cm]{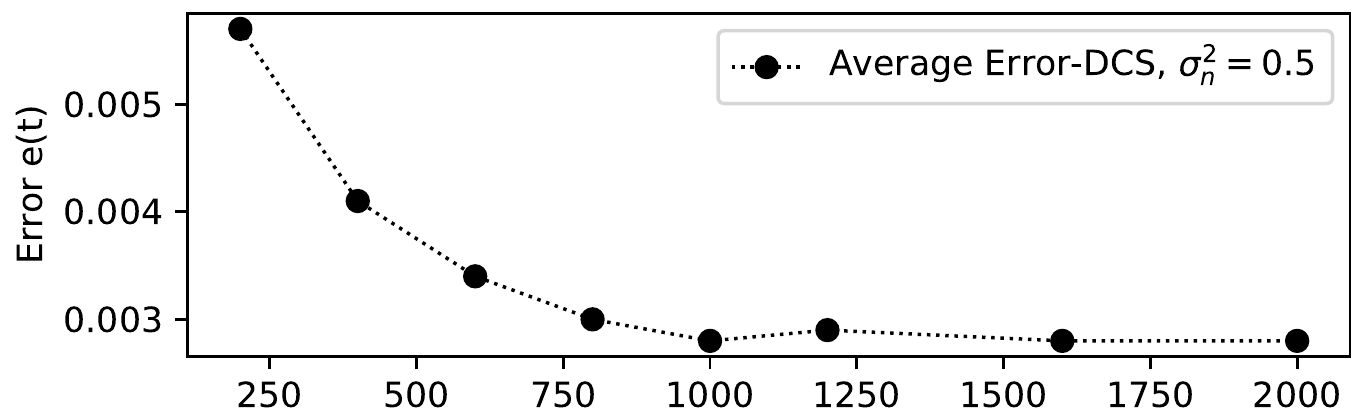} 
\begin{center}
\small {Number of Iterations}\\
(a)
\end{center}

\includegraphics[width=8.5cm]{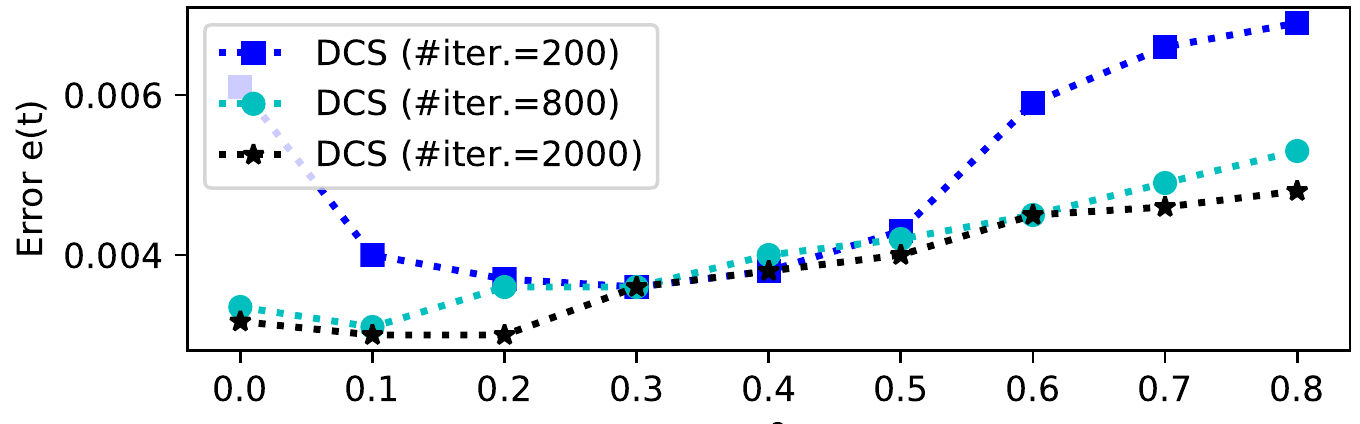}
\begin{center}
\small{$\sigma^2_n$}\\
(b)
\end{center}

\includegraphics[width=8.5cm]{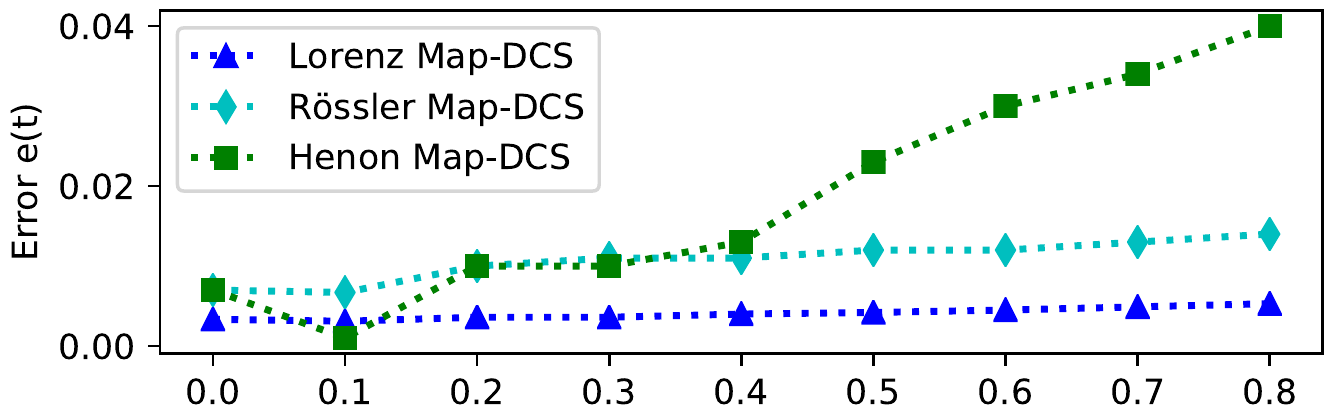}
\begin{center}
\small{$\sigma^2_n$}\\
(c)
\end{center}

\caption{\small{(a) DCS convergence. 
(b) Effect of increasing the number of  iterations. (c) A comparison among the noise performance of the Lorenz, Rössler, and Henon maps. }}

\end{figure}
\subsection{Synchronization Error Comparison}
The following simulations deal with the output signals of the response system. Fig. 8 show a comparison between the output of the traditional Lorenz coupled system in (3) and the proposed DCS. Fig. 8 (a) shows the master signal $z(t)$ and reconstructed attractor $z_r(t)$ using the traditional system. The parameters in the transmitter and receiver of the conventional system are the same.  We assume that in the receiver, the initial state of $(x_0)$ is a random number in the range [0, 0.1]. Fig. 8 (b) shows the reconstructed attractor $z_r(t)$ by the DCS. The amplitude of the synchronization error is very small because of the noise reduction and  initialal state estimation in DCS, and the synchronization is more persistent over time.
\begin{figure}

\includegraphics[width=8.5cm]{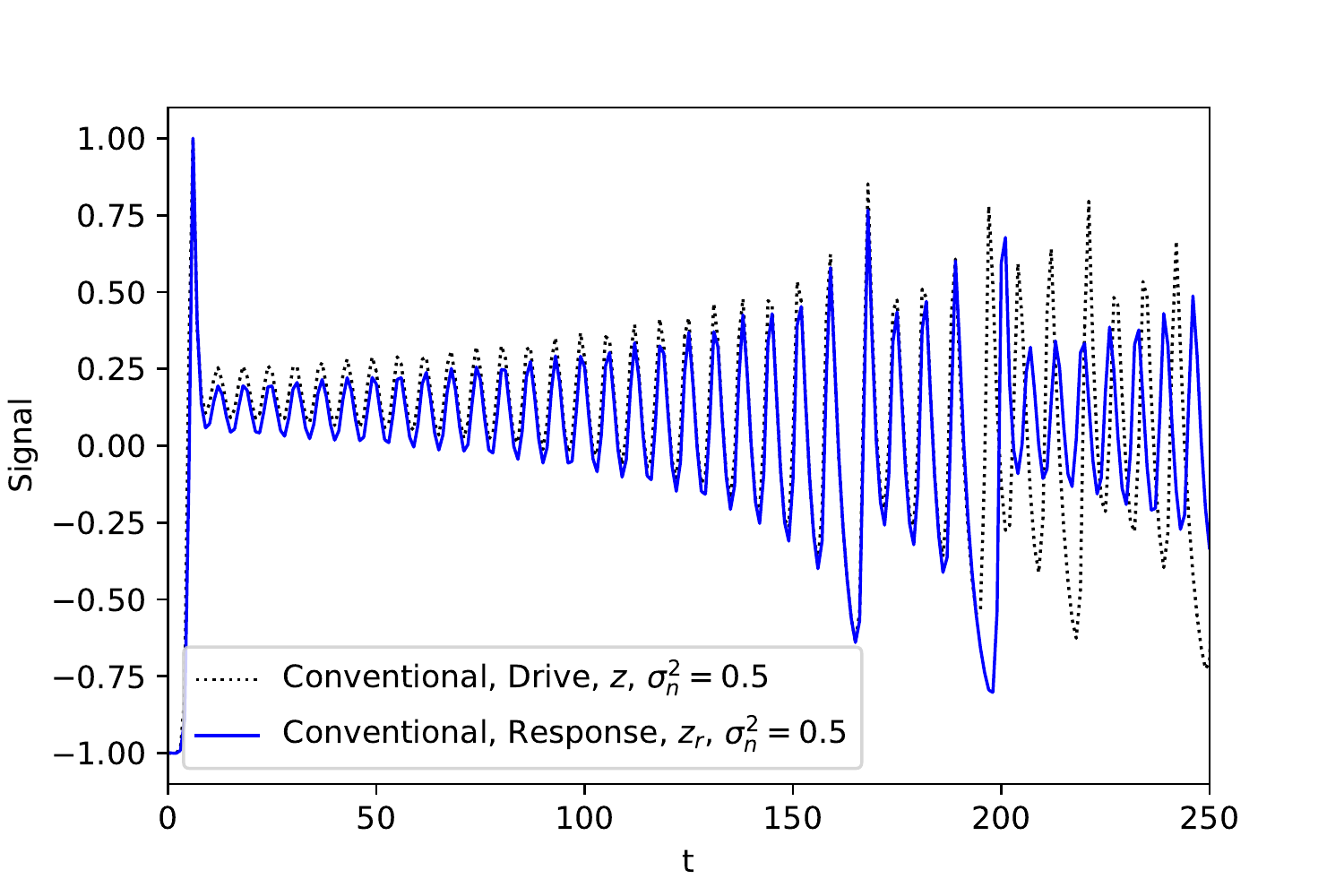} 
\begin{center}
~~~~~~~~(a)
\end{center}

\includegraphics[width=8.5cm]{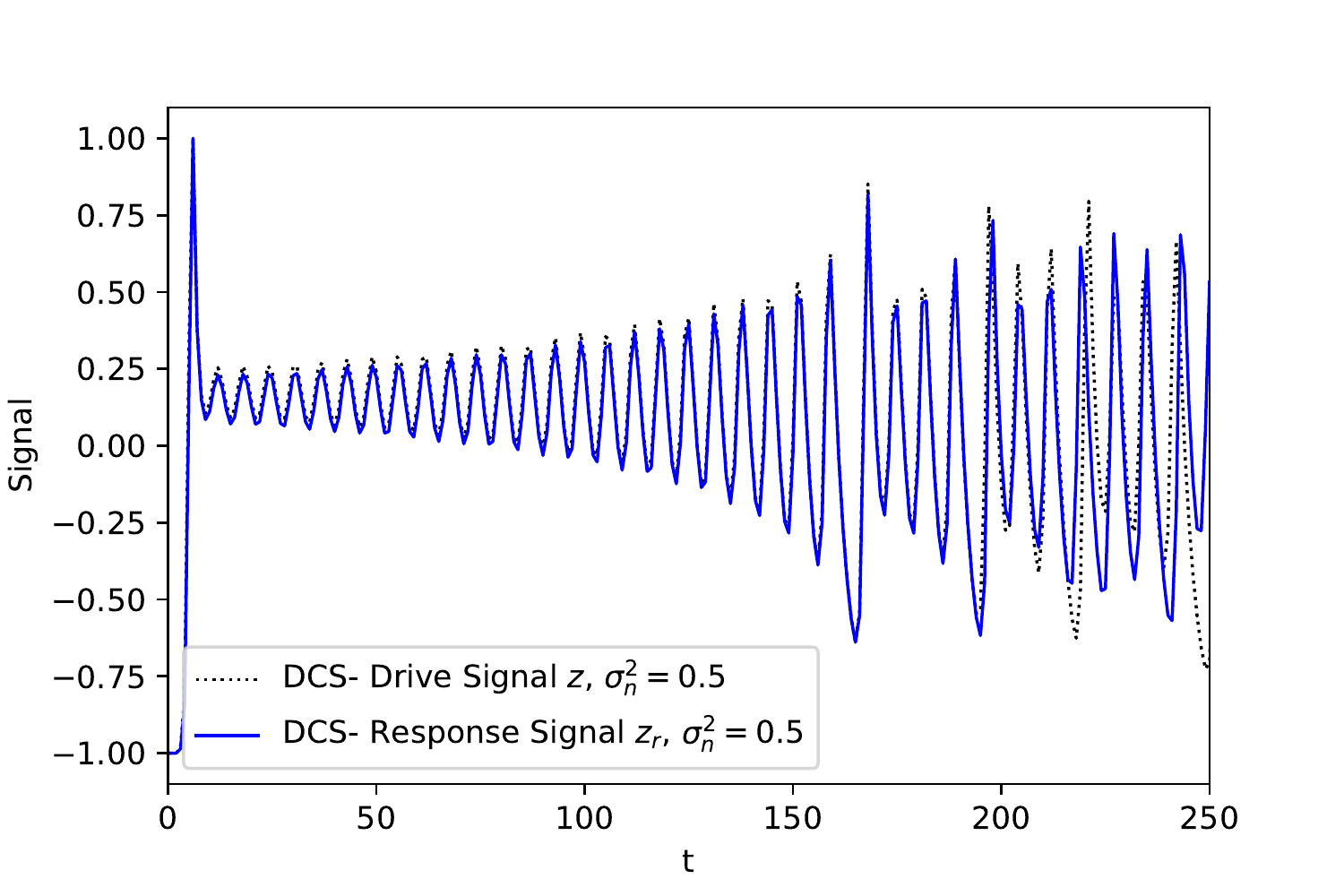}
\begin{center}
~~~~~~~~(b)
\end{center}
\caption{\small{Master signal z(t) and reconstructed signal in (a) conventional system (b) DCS.}}

\end{figure}\par
 Fig. 9 presents $z(t)-z_r(t)$ as the synchronization error criterion. To make the visual comparison easier, error signals are plotted over a smaller time span. In comparison with the traditional coupled Lorenz system and the RNN-based synchronization system, the DCS model conserves the dynamics of the master signal while it has a very small error amplitude.
 \begin{figure}[htp]
    \centering \includegraphics[width=8.6cm]{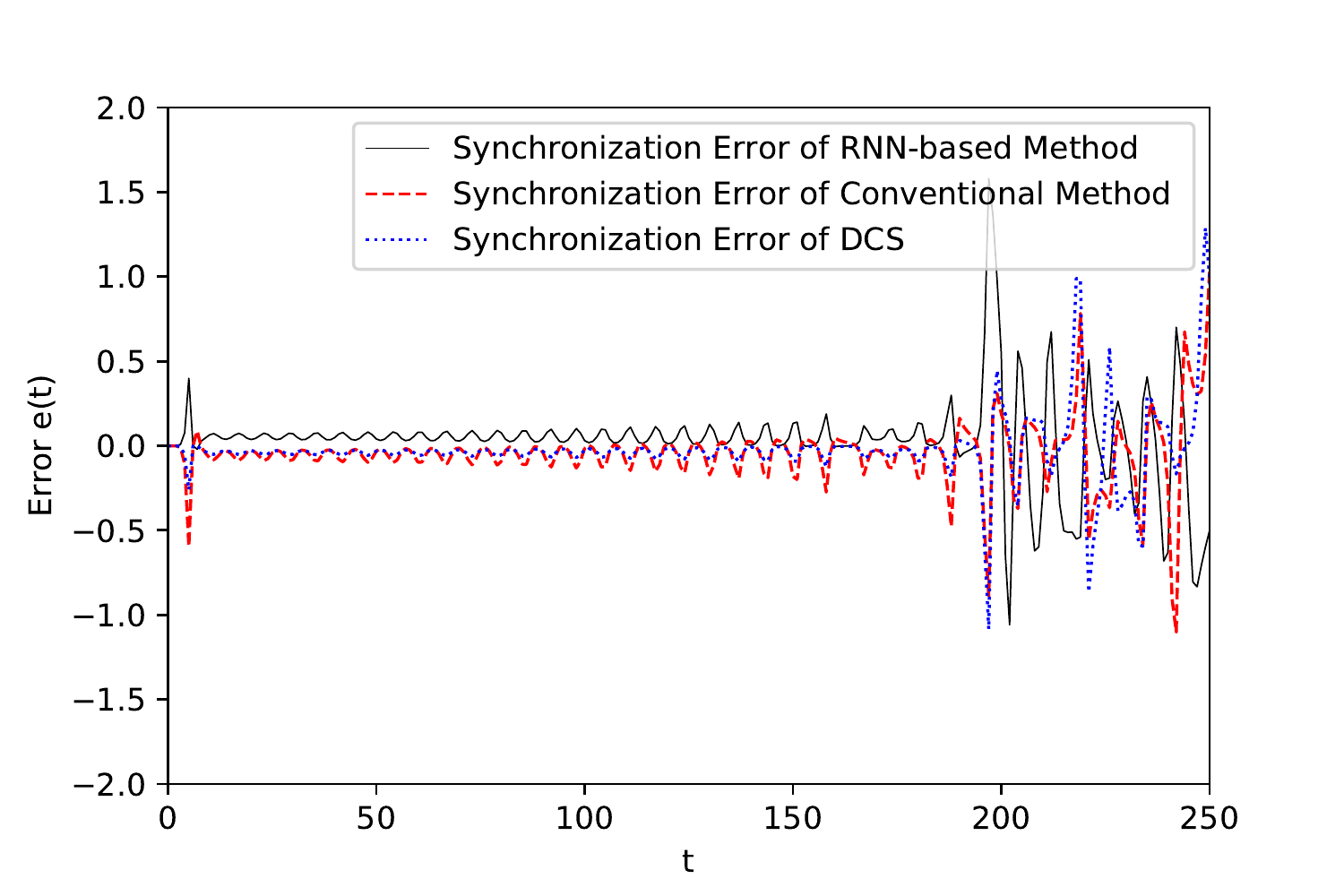}
    \centering
    \caption{\small { Comparison between synchronization error of the different systems for{} $\sigma_n^2=0.5$.}}
    \label{fig:Picture1}
\end{figure}
Because the visual comparison may not be accurate, in another experiment we measured the average synchronization error for the above mentioned span $(t=250)$ and for different values of the noise power. The results are shown in the Table III. Observe that the DCS method is always superior to the other two methods.
\begin{table}[!t]

\renewcommand{\arraystretch}{1.2}

\caption{\small{Average synchronization error of different methods for different noise values.}}

\centering
\begin{tabular}{|c||c||c||c|}
\hline
 &RNN-based Method&Conventional Method&DCS\\
\hline
$\sigma_n^2=0.1$ &0.05101&0.07180& 0.02791\\
\hline
$\sigma_n^2=0.2$ & 0.05111
&0.07181&0.02851\\
\hline
$\sigma_n^2=0.3$ & 0.04428&0.07171&0.02779\\
\hline
$\sigma_n^2=0.4$ & 0.05193 &0.07192&0.02780\\
\hline
$\sigma_n^2=0.5$ &0.05195 &0.07195&0.02781\\
\hline
$\sigma_n^2=0.6 $ &0.05501 &0.07196&0.02781\\
\hline
$\sigma_n^2=0.7$ & 0.06333 &0.07197&0.02782\\
\hline
\end{tabular}
\end{table}
More exactly, consider a scenario, where a chaotic signal with 1024 samples generated, and portions of the signal with specified lengths are randomly selected for transmission and performance evaluation. We repeated this experiment ten times for each determined length and calculated the average synchronization error. \par
As shown in Fig. 10, the DCS system has a competitive performance with the RNN-based system and even works better for signals with $T>250$. Furthermore, in this figure, both proposed systems have been compared with the evolutionary de-noising method presented in [75]. To make a fair comparison, GA-based initial condition estimation is used for all of them, according to the values in table 2. because of the benefits of neural networks, both of the proposed systems are more robust than the evolutionary de-noising method that only uses initial condition estimation, specifically when $T>250$.
It is important to note that we have deliberately selected large noise and high perturbation of the initial values in this experiment so that the performance of the various methods is visually distinguishable. Adjusting parameters like the initial population and mutation rate in GA can counteract these damaging effects and eliminate synchronization errors.\par

 \begin{figure}[htp]
    \centering \includegraphics[width=8.5cm]{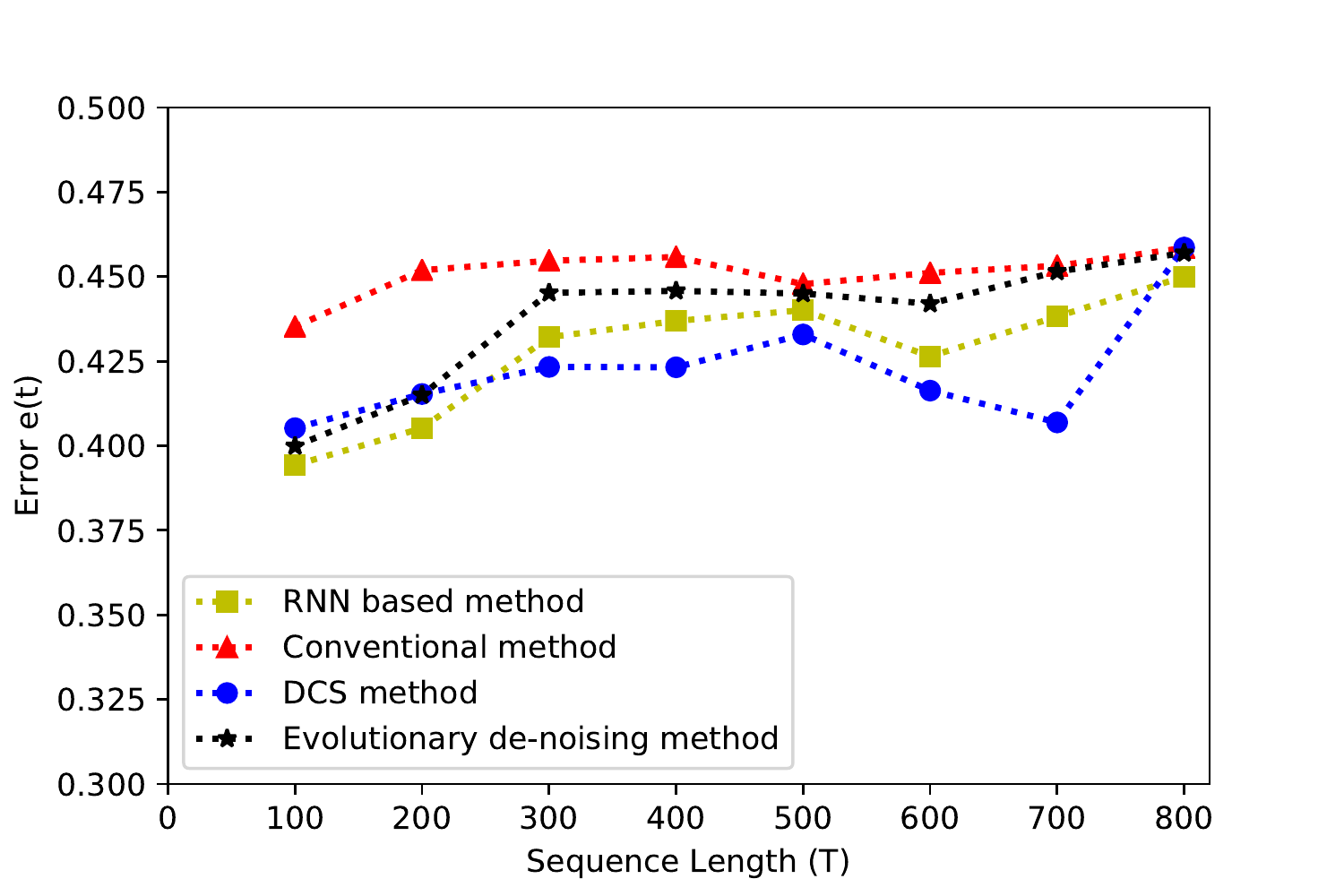}
    \centering
    \caption{\small{Comparison between synchronization error of the different systems for{} $\sigma_n^2=0.5$.}}
    \label{fig:Picture1}
\end{figure}
\section{Conclusions}
This paper presents the ﬁrst attempt of using DL in chaos-based communication systems with providing an excellent synchronization between two coupled Lorenz systems. The results showed that the DCS reduces the synchronization error compared to the traditional systems and RNN-based approach. Moreover, the proposed DCS is easy to implement since it does not require training on large data sets. Thus, it can be used in coherent wireless communication scenarios where users or mobile units are generally hard to train. Practical implementations of DCS-based applications e.g., chaos-based CDMA systems and low latency communications is an interesting topic for future researches.
\par

\section*{Acknowledgment}
We are grateful to the Tier2 Canada research chair entitled
\textbf{ ‘Towardsa Novel and Intelligent Framework for the Next generations of IoT Networks’} for support this project.

\begin{IEEEbiography}[{\includegraphics[width=1in,height=1.25in,clip,keepaspectratio]{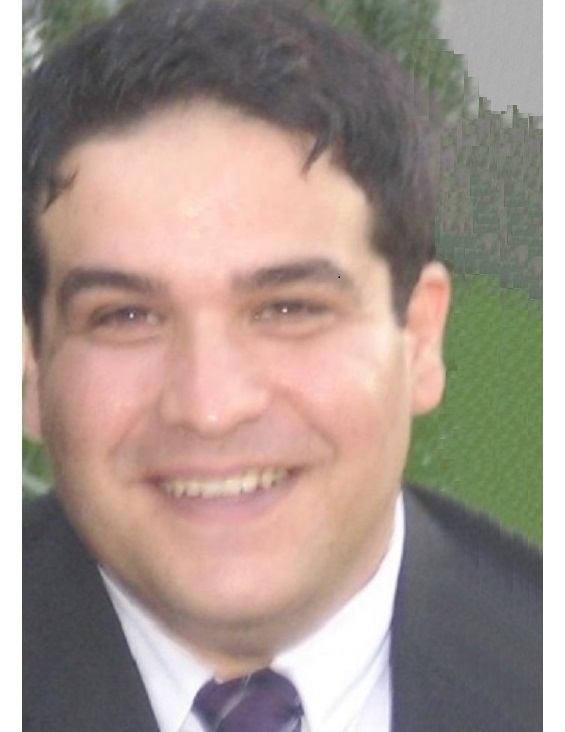}}]{MAJID MOBINI}received his M.Sc. degree in communications engineering with the first rank from the Department of Electrical and Computer Engineering, Amirkabir University of Technology (Tehran Polytechnic), Tehran, Iran, in 2013, and the Ph.D. degree in communications engineering from the Babol Noshirvani University of Technology, in 2019. He was a Visiting Scholar with Cyberspace Research Institute, Shahid Beheshti University, Tehran, Iran, from 2013 to 2014. He is currently a Research Associate with the Department of Electrical and Computer Engineering, Babol Noshirvani University of Technology. His current research interests include chaos-based wireless Communication, optimization, artiﬁcial intelligence, Internet of Things, cyber-physical systems, machine vision, biomedical instruments, and interdisciplinary researches. He is a Reviewer of the IEEE ACCESS and the Journal of Soft Computing and Information Theory.  
\end{IEEEbiography}
\vskip 0pt plus -1fil
\begin{IEEEbiography}[{\includegraphics[width=1in,height=1.25in,clip,keepaspectratio]{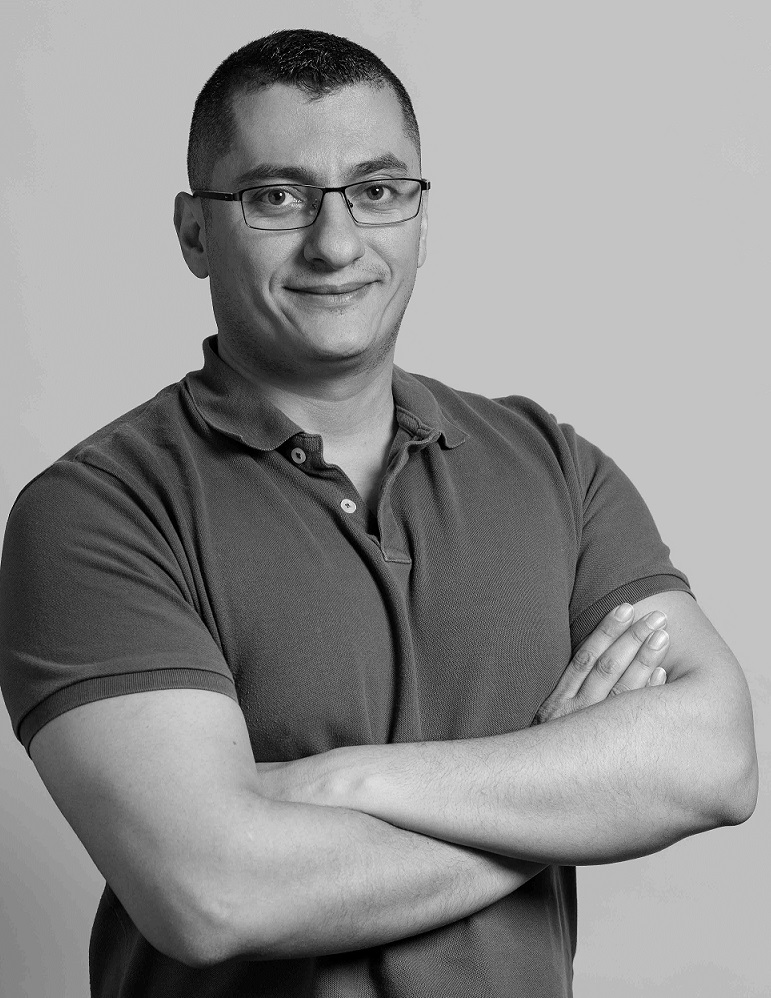}}]{GEORGES KADDOUM} received the Bachelor’s degree in electrical engineering from the \'Ecole Nationale Supérieure de Techniques Avancées (ENSTA Bretagne), Brest, France, and the M.S. degree in telecommunications and signal processing(circuits, systems, and signal processing) from the Universit\'e de Bretagne Occidentale and Telecom Bretagne (ENSTB), Brest, in 2005 and the Ph.D. degree (with honors) in signal processing and telecommunications from the National Institute of Applied Sciences (INSA), University of Toulouse, Toulouse, France, in 2009. He is currently an Associate Professor and Tier 2 Canada Research Chair with the \'Ecole de Technologie Sup\'erieure (\'ETS), Universit\'e du Qu\'ebec, Montr\'eal, Canada. In 2014, he was awarded the \'ETS Research Chair in physical-layer security for wireless networks.  Since 2010, he has been a Scientific Consultant in the field of space and wireless telecommunications for several US and Canadian companies. He has published over 200+ journal and conference papers and has two pending patents. His recent research activities cover mobile communication systems, modulations, security, and space communications and navigation. Dr. Kaddoum received the Best Papers Awards at the 2014 IEEE International Conference on Wireless and Mobile Computing, Networking, Communications (WIMOB), with three coauthors, and at the 2017 IEEE International Symposium on Personal Indoor and Mobile Radio Communications (PIMRC), with four coauthors. Moreover, he received IEEE Transactions on Communications Exemplary Reviewer Award for the year 2015, 2017, 2019. In addition, he received the research excellence award of the Universit\'e du Qu\'ebec in the year 2018. In the year 2019, he received the research excellence award from the \'ETS in recognition of his outstanding research outcomes. Prof. Kaddoum is currently serving as an Associate Editor for IEEE Transactions on Information Forensics and Security, and IEEE Communications Letters.
\end{IEEEbiography}

\end{document}